
\magnification=1200
\def\build#1_#2^#3{\mathrel{\mathop{\kern 0pt#1}\limits_{#2}^{#3}}}
\catcode `@=11
\baselineskip 18pt
\hsize 16truecm
\vsize 24truecm

\def\n{\noindent}
\def\m{\medskip}
\def\b{\bigskip}

\font\twelve=cmbx10 at 13pt
\font\eightrm=cmr8
\font\sevenrm=cmr7

\def\lesssim{\mathrel{\mathpalette\vereq<}}
\def\vereq#1#2{\lower3pt\vbox{\baselineskip1.5pt \lineskip1.5pt
\ialign{$\m@th#1\hfill##\hfil$\crcr#2\crcr\sim\crcr}}}

\def\tv{\tvi\vrule}

{
\centerline{\twelve CENTRE DE PHYSIQUE THEORIQUE}
\centerline{\twelve CNRS - Luminy, Case 907}
\centerline{\twelve 13288 Marseille Cedex}

\vskip 4truecm

\centerline{{\twelve CHIRAL PERTURBATION
THEORY}\footnote{$^{\displaystyle\star}$}{\eightrm Invited talk
at the HADRON'93 Conference Como, Italy June 1993.}}

\bigskip

\centerline{\bf Eduardo de Rafael}

\vskip 3truecm

{\leftskip=1cm
\rightskip=1cm

\centerline{\bf Abstract}
\medskip
The basic ideas and some recent developments of the chiral
perturbation theory approach to hadron dynamics at low-energies
are reviewed.

\vskip 6truecm

\n December 1993

\n CPT-93/P.2967

\par}

\footline={}
\vfill\eject
}

\pageno=1

Chiral perturbation theory ($\chi$PT) is the effective field theory
of quantum chromodynamics (QCD) at low-energies. In this talk I
shall first give a brief review of the basic ideas of the
$\chi$PT-approach to hadron dynamics at low-energies. Then I shall
review some of the phenomenological applications of $\chi$PT with
emphasis on recent developpements. I also want to discuss the {\it
limitations} of the $\chi$PT-approach. This will bring me to the
related question of how to derive a low-energy effective Lagrangian
{}from QCD and to review recent work in this direction.

\b

{\bf 1.} The principle of the effective field theory approach to the
description of low-energy phenomena can best be illustrated with
a simple example~: the effect of hadronic vacuum polarization at
very low energies. At very low momentum transfer $q^2$ --- low
compared to the masses of the virtual hadronic pair creation --- we
can expand the photon self-energy $\Pi(q^2)$ in powers of
momenta~:
$$\Pi(q^2)=(\Pi(0)=0)+\left.{\partial\Pi\over\partial
q^2}\right|_{q^2=0}q^2+\dots,\eqno(1)$$
and approximate the hadronic photon-self energy by its slope at
the origin. (The fact that $\Pi(0)=0$ is due to the electric charge
renormalization.) This approximation is best described by the
effective Lagrangian which results from integrating out the
hadronic degrees of freedom of the underlying theory in the
presence of the electromagnetic interactions. The form of the
resulting effective Lagrangian of quantum electrodynamics (QED)
can be written down using gauge invariance alone~:
$${\cal L}^{QED}_{eff}=-{1\over
4}\left\{F^{\mu\nu}(x)F_{\mu\nu}(x)-{1\over
\Lambda^2}\partial^{\lambda}F^{\mu\nu}(x)\partial_{\lambda}
F_{\mu\nu}(x)+\dots\right\}.\eqno(2)$$
The effective local interaction of dimension six which appears,
describes in a universal way the physics due to a non-zero slope of
the hadronic photon self-energy. The value of the constant
$\Lambda^2$ is not fixed by arguments of symmetry alone. However,
once it is determined from one observable --- say the $g-2$ of the
muon --- we have well defined predictions for many other
observables like e.g. the lamb-shift~; electron-electron scattering
etc.~[1]. Only if we know the dynamics of the underlying theory can
we attempt to a calculation of $\Lambda^2$. For example, we can
easily calculate the contribution to $\Lambda^2$ from the
electromagnetic interactions due to the heavy quarks $c,\ b$ and
$t$. Their contribution can be well approximated by their lowest
order electromagnetic couplings and, with neglect of gluonic
corrections which are small at the heavy quark mass scale, we get
($e_i$ is the electric charge of the quark $i=c,\ b,\ t$ in $e$
units~; $N_c$ the number of QCD-colours)
$${1\over \Lambda^2_i}\simeq{\alpha\over \pi}e^2_i{N_c\over
15}{1\over M^2_i}\ \raise 2pt\hbox{,}\eqno(3)$$
explicitly showing, in this case, the decoupling of heavy quark
effects in low energy QED-physics.

\m

This simple example illustrates the three basic ingredients of the
effective Lagrangian approach~:

\parindent 1truecm
\item{\hbox to\parindent{\enskip \hfill i) \hfill}} The structure of
the local interaction is fixed by the symmetry properties of the
underlying theory~; in this case gauge invariance.

\item{\hbox to\parindent{\enskip \hfill ii) \hfill}} The domain of
validity of the effective approach is restricted to processes
governed by values of momenta smaller than a characteristic
scale~; in this case the hadronic mass threshold corresponding to
the quark-f $\! $lavour which has been integrated out.

\item{\hbox to\parindent{\enskip \hfill iii) \hfill}} The coupling
constants of the effective Lagrangian, like $\Lambda^{-2}$ in our
case, are not fixed by arguments of symmetry alone. Only if we know
the details of the underlying dynamics can we calculate them~; as
we have illustrated in the case of the heavy quark contributions to
$\Lambda^{-2}$ in (3).

\b

{\bf 2.} In the limit where the masses of the light quarks $u,\ d$
and $s$ are set to zero, the QCD Lagrangian is invariant under
rotations $(V_L,\ V_R)$ of the left-and right-handed quark triplets
$q_L\equiv {1-\gamma_5\over 2}q$ and $q_R\equiv
{1+\gamma_5\over 2}q$~; $q=u,\ d,\ s$. These rotations generate
the so scaled chiral-$SU(3)$ group~: $SU(3)_L\times SU(3)_R$. At
the level of the hadronic spectrum, this symmetry of the QCD
Lagrangian is however spontaneously broken down to the diagonal
$SU(3)_V,\ V=L+R$. The reduced invariance is the famous
$SU(3)$-symmetry of the Eightfold Way~[2]. This pattern of
spontaneously broken symmetry implies specific constraints on the
dynamics of the strong interactions between the low-lying
pseudoscalar states $(\pi,\ K,\ \eta)$, which are the massless
Goldstone bosons associated to the ``broken'' chiral generators. As a
result of the spontaneous symmetry breaking, there appears a
mass-gap in the hadronic spectrum between the ground state of the
octet of $0^-$-pseudoscalars and the lowest hadronic states which
become massive in the chiral limit $m_u=m_d=m_s=0$~; i.e., the
octet of $1^-$-vector-meson states and the octet of $1^+$
axial-vector-meson states. The basic idea of the
$\chi$PT-approach is that in order to describe the physics at
energies within this gap region, it may be more useful to formulate
the strong interactions of the low-lying pseudoscalar particles in
terms of an effective low-energy Lagrangian of QCD, with the
octet of Goldstone fields ($\overrightarrow{\lambda}$ are the
eight $3\times 3$ Gell-Mann matrices)
$$\phi(x)={\overrightarrow{\lambda}\over
\sqrt{2}}\cdot\overrightarrow{\varphi}(x)=
\normalbaselineskip=18pt\pmatrix{\pi^{\circ}/\sqrt{2}+
\eta/\sqrt{6}&\hfill\pi^+\hfill &\hfill K^+\hfill\cr
\hfill\pi^-\hfill &-\pi^{\circ}/\sqrt{2}+\eta/\sqrt{6}&
\hfill K^{\circ}\hfill \cr
\hfill K^- \hfill &\hfill\overline K^{\circ} \hfill &-2\eta/\sqrt{6}
}\eqno(4)$$
as explicit degrees of freedom, rather than in terms of the quark
and gluon fields of the usual QCD Lagrangian.

The most general effective Lagrangian, compatible with the
symmetry pattern des\-cribed above is a non-linear Lagrangian
with the octet of fields $\overrightarrow{\varphi}(x)$ in (4)
collected in a unitary $3\times 3$ matrix ${\cal U}(x)$ with
$\det {\cal U}=1$. Under chiral rotations $(V_L,\ V_R)$ the matrix
${\cal U}$ is chosen to transform linearly
$${\cal U}\to V_R\ {\cal U}\ V^{\dagger}_L.\eqno(5)$$ The
effective Lagrangian we look for has to be then a sum of chiraly
invariant terms with increasing number of derivatives of ${\cal
U}$. For example, to lowest order in the number of derivatives, only
one independent term can be constructed which is invariant under
$(V_L,\ V_R)$ transformations~:
$${\cal L}_{eff}={1\over 4}f^2_{\pi}tr \partial_{\mu}{\cal
U}(x)\partial_{\mu}{\cal U}^{\dagger}(x),\eqno(6)$$ where the
normalization is fixed in such a way that the axial-current deduced
{}from this Lagrangian induces the experimentally observed
$\pi\to\mu\nu$ transition. An explicit representation of ${\cal U}$
is
$${\cal U}(x)=\exp\left(-i{1\over
f_{\pi}}\overrightarrow{\lambda}\cdot
\overrightarrow{\varphi}(x)\right)\ ;\eqno(7a)$$
and
$$f_{\pi}=93.2\ MeV.\eqno(7b)$$
Because of the non-linearity in ${\varphi}$, processes with
different number of pseudoscalar mesons are then related. These
are the successful current-algebra relations~[3] of the 60's which
the effective Lagrangian above incorporates in a compact way~[4].

It is useful to promote the global chiral-$SU(3)$ symmetry to a
local $SU(3)_L\times SU(3)_R\ $ gauge symmetry. This can be
accomplished by adding appropriate quark bi\-li\-near couplings
with external field sources to the usual QCD-Lagrangian ${\cal
L}^{\circ}_{QCD}$~; i.e.,
$${\cal L}_{QCD}(x)={\cal L}^{\circ}_{QCD}(x)+\overline
q\gamma^{\mu}(v_{\mu}+\gamma_5 a_{\mu})q-\overline
q(s-i\gamma_5 p)q.\eqno(8)$$ The external field sources
$v_{\mu},\ a_{\mu},\ s$ and $p$ are Hermitian $3\times 3$
matrices in flavour and colour singlets. In the presence of these
external field sources, the possible terms in
${\cal L}_{eff}$ with the lowest chiral dimension, i.e., $O(p^2)$ are
$${\cal L}_{eff}={1\over 4}f^2_{\pi}\left\{tr D_{\mu}{\cal
U}D^{\mu}{\cal U}^+ +tr(\chi{\cal U}^+ +{\cal
U}\chi^+)\right\},\eqno(9)$$
where
$D_{\mu}$ denotes the covariant derivative
$$D_{\mu}{\cal U}=\partial_{\mu}{\cal U}-i(v_{\mu}+a_{\mu}){\cal
U}+i{\cal U}(v_{\mu}-a_{\mu})\eqno(10)$$ and
$$\chi=2B(s(x)+i p(x)),\eqno(11)$$ with $B$ a constant, which like
$f_{\pi}$, is not fixed by symmetry requirements alone. Once
special directions in flavour space (like the ones selected by the
electroweak Standard Model couplings) are fixed for the external
fields, the chiral symmetry is then explicitly broken. In particular,
the choice
$$s+i p={\cal M}=\hbox{diag}\ (m_u,\ m_d,\ m_s)\eqno(12)$$
takes into
account the explicit breaking due to the quark masses in the
underlying QCD Lagrangian. In the conventional picture of chiral
symmetry breaking, the constant $B$ is
related to the light quark condensate
$$\left<0|\overline
q^jq^i|0\right>=-f^2_{\pi}B\delta_{ij},\eqno(13)$$
and the relation
between the physical pseudoscalar masses and the quark masses,
to lowest order in the chiral expansion, is then fixed by identifying
quadratic terms in ${\varphi}$ in the expansion of the second term
in (9), with the result
$$\chi=\normalbaselineskip=18pt\pmatrix{
m^2_{\pi^+}+M^2_{K^+}-M^2_{K^0} &\hfill 0\hfill &\hfill 0\hfill\cr
\hfill 0\hfill & m^2_{\pi^+}-M^2_{K^+}+M^2_{K^0} &\hfill 0\hfill\cr
\hfill 0\hfill &\hfill 0\hfill & -m^2_{\pi^+}+M^2_{K^+}+M^2_{K^0}
}.\eqno(14)$$

The effective Lagrangian (9) describes physical $S$-matrix
amplitudes to order $O(p^2)$ in momenta~:
$$A(p_1,\ p_2,\ \dots)=\Sigma a_{ij}p_i\cdot p_j+O(p^4),$$
and
predicts all the possible $a_{ij}$ couplings. The apparent absence
of terms of $O(p^{\circ})$ is a result of the chiral invariance of the
underlying QCD-theory\footnote{$^1$}{\eightrm Terms of
$O(p^{\circ})$ do appear, however, in the presence of virtual
electromagnetic interactions.}. There are three sources of possible
contributions to $O(p^4)$~:

\parindent 1truecm
\item{\hbox to\parindent{\enskip \hfill a) \hfill}} Tree level
amplitudes from the local $O(p^4)$ effective couplings. (More on
that soon.)

\item{\hbox to\parindent{\enskip \hfill b) \hfill}} One-loop
Feynman diagrams generated by the lowest order effective
Lagrangian in (9). Loops have to be taken into account to guarantee
$S$-matrix unitarity at the level of approximation one is working
with.

\item{\hbox to\parindent{\enskip \hfill c) \hfill}} Amplitudes
generated by the presence of the chiral anomaly. The corresponding
effective Lagrangian is known from the work of Bardeen~[5] and
Wess and Zumi\-no~[6]. (See also Witten~[7].) A typical process
fully accounted by this type of contribution is the decay
$\pi^{\circ}\to\gamma\gamma$, which is forbidden to
$O(p^2)$.

\m

The identification of all the independent local terms of $O(p^4)$,
invariant under pa\-ri\-ty, charge-conjugation and local
chiral-$SU(3)$ transformations~; as well as the phenomenological
determination of their corresponding ten physical coupling
constants
$L_i,\ i=1,\ 2,\
\dots,\ 10$ has been done by Gasser and Leutwyler in a series of
seminal papers~[8]. We follow their notation~:
$$\eqalignno{ {\cal L}^{(4)}_{eff} &=L_1\left(tr D_{\mu}{\cal
U}^{\dagger}D^{\mu}{\cal U}\right)^2+L_2 tr D_{\mu} {\cal
U}^{\dagger}D_{\nu}{\cal U}tr D^{\mu}{\cal U}^{\dagger}D^{\nu}{\cal
U}\cr &+L_3 tr D_{\mu}{\cal U}^{\dagger}D^{\mu}{\cal
U}D_{\nu}{\cal U}^{\dagger}D^{\nu}{\cal U}\cr &+L_4 tr
D_{\mu}{\cal U}^{\dagger}D^{\mu}{\cal U}
tr\left(\chi^{\dagger}{\cal U}+{\cal U}^{\dagger}\chi\right)+L_5 tr
D_{\mu}{\cal U}^{\dagger}D^{\mu}{\cal U}\left(\chi^{\dagger}{\cal
U}+{\cal U}^{\dagger}\chi\right)\cr &+L_6\left[
tr\left(\chi^{\dagger}{\cal U}+{\cal
U}^{\dagger}\chi\right)\right]^2+L_7\left[
tr\left(\chi^{\dagger}{\cal U}-{\cal
U}^{\dagger}\chi\right)\right]^2+L_8 tr\left({\cal
U}\chi^{\dagger}{\cal U}\chi^{\dagger} +{\cal U}^{\dagger}\chi\
{\cal U}^{\dagger}\chi\right)\cr &+iL_9 tr\left(F^{\mu\nu}_R
D_{\mu}{\cal U}D_{\nu}{\cal U}^{\dagger} +F^{\mu\nu}_L
D_{\mu}{\cal U}^{\dagger}D_{\nu}{\cal U}\right)+L_{10} tr {\cal
U}^{\dagger}F^{\mu\nu}_R{\cal U}F_{L\mu\nu}\cr &+H_1 tr
\left(F^{\mu\nu}_R F_{R\mu\nu}+F^{\mu\nu}_L
F_{L\mu\nu}\right)+H_2 tr
\chi^{\dagger}\chi.&(15) }$$
Here $F^{\mu\nu}_L(x)$ and
$F^{\mu\nu}_R(x)$ are the non-abelian field-strength tensors
associated to the external left
$(\ell_{\mu}=v_{\mu}-a_{\mu})$ and right
$(r_{\mu}=v_{\mu}+a_{\mu})$ field sources. Notice that the last
two terms in~(15) involve external fields only.

The coupling constants $L_i$ are dimensionless. If the chiral
expansion makes sense, their expected order of magnitude should
be
$$L_i\simeq{1\over 4}f^2_{\pi}{1\over
\Lambda^2_{\chi}}\ \raise 2pt\hbox{,}\eqno(16)$$
with $\Lambda_{\chi}$ the scale of
spontaneous chiral symmetry breaking which we expect to be of
the order of the masses of the lowest $1^-,\ 1^+$ hadron states i.e.,
$$M_{\rho}(770\ MeV)\lesssim\Lambda_{\chi}\lesssim M_A(1260\
MeV).\eqno(17a)$$
This corresponds to
$$10^{-3}\lesssim L_i \lesssim 5\cdot 10^{-3}.\eqno(17b)$$ In
Table 1, I have collected the most recent phenomenological
determination of the
$L_i$'s~[9]. These values correspond to the renormalized couplings
at the scale of the $\rho$-mass. The corresponding values at
another scale $\mu$ are given by the one-loop renormalization
group equation
$$L_i(\mu)=L_i(M_{\rho})+{\Gamma_i\over
16\pi^2}\log{M_{\rho}\over\mu}\ \raise 2pt\hbox{,}\eqno(18)$$
with $\Gamma_i$ the numbers collected in the last column of
Table 1. I have also indicated in this table the sources of the
determination of the
$L_i$'s. The agreement with the ``expected'' order of magnitude in
(17) is quite reasonable and justifies, a posteriori, the use of an
effective Lagrangian description for low-energy hadron physics.

\b

{\bf 3.} During the last ten years, there has been a wealth of
applications of $\chi$PT to physical processes. For recent review
articles see e.g. refs.~[10, 11, 12]. I shall illustrate this
acti\-vity with a few relatively recent examples~:

\m

\n {\bf 3a. The $\pi -\pi$ phase shift difference
$\delta^{\circ}_{\circ}(M^2_K)-\delta^2_{\circ}(M^2_K)$.}

You may remember that it is this difference which governs the
phase of the CP-violation parameter $\epsilon '$. Gasser and
Mei\ss ner have performed an $O(p^4)$ calculation of this phase
difference with the result~[13]
$$\delta^{I=0}_{J=0}(M^2_K)-\delta^{I=2}_{J=0}(M^2_K)=45^{\circ}
\pm 6^{\circ}.\eqno(19)$$
The corresponding result to $O(p^2)$ was
$\delta^{\circ}_{\circ}-\delta^2_{\circ}=37^{\circ}$.

\m

\n {\bf 3b. Semileptonic $K$-decays.}

There has been a lot of work in this sector, essentially motivated
by the physics program of the forthcoming DA$\phi$NE facility at
Frascati. Let me also remind you that the most precise
determination of $V_{us}$~:
$$|V_{us}|=0.2196\pm 0.0023,\eqno(20)$$ comes from the analysis
of $K_{\ell_3}$-data within the framework of $\chi$PT~[14].

In the large $N_c$ limit of QCD ($N_c$ is the number of colors)~:
$L_2=2L_1$. The $K_{\ell_4}$-decays offer a unique test of the
validity of this approximation. The analysis to $O(p^4)$ made in
refs.~[15, 16] gives the result
$${(L_2-2L_1)\over L_3}=-0.19^{+0.16}_{-0.27}\ .\eqno(21)$$

\b

The vector form factor of $K_{\ell_4}$ at $q^2=0$ is predicted by
the Wess-Zumino anoma\-lous term of the effective chiral
Lagrangian I already mentioned, with the result~[9]
$$H(0)=-{\sqrt{2}M^3_K\over 8\pi^2 f^3_{\pi}}=-2.7\ .\eqno(22)$$
Experimentally~[17]
$$H_{\exp}=-2.68\pm 0.68\ .\eqno(23)$$

\m

\n {\bf 3c. $\eta$-decays and
$\gamma\gamma\to\pi^{\circ}\pi^{\circ}$.}

Here we enter a class of processes which demand rather detailed
knowledge of $\chi$PT beyond the $O(p^4)$. The full two-loop
calculation of the amplitude
$\gamma\gamma\to\pi^{\circ}\pi^{\circ}$ has been recently
reported~[18]. This calculation, which is quite a technical
performance, when complemented with estimates of the relevant
$O(p^6)$ tree level couplings leads to a prediction for the
low-energy behaviour of the integrated cross-section in good
agreement with data from the Crystal Ball collaboration~[19].

\b

{\bf 4.} There has also been quite a lot of progress in understanding
the r\^ole of resonances in the $\chi$PT-approach. It has been
shown that the values of the $L_i$-constants are practically
saturated by the lowest resonance exchanges between
pseudoscalar particles~; and  particularly by vector-exchange
whenever vector mesons can contribute~[20, 21].

The specific form of an effective chiral invariant Lagrangian
describing the couplings of vector and axial-vector particles to the
Goldstone modes is not uniquely fixed by chiral symmetry
requirements alone. When the vector particles are integrated out,
different field theory descriptions may lead to different
predictions for the $L_i$'s. It has been shown however that if a
few QCD short-distance constraints are imposed, the ambiguities
of different formulations are then removed~[22]. The most compact
effective Lagrangian formulation, which is compatible with the
short-distance constraints, has two parameters~: $f_{\pi}$ and
$M_V$. When the vector and axial-vector fields are integrated out,
it leads to specifie predictions for five of the $L_i$ constants in
good agreement with experiment.

We can conclude that the old phenomenological concept of vector
meson dominance $(VMD)$~[23] can now be formulated in a way
compatible with the chiral symmetry properties of QCD as well as
its dynamical short-distance behaviour. Further extensions of these
results have been developed more recently in refs.~[24] and ~[25].

\b

{\bf 5.} The $\chi$PT-approach has also been successfully applied
to the sector of the hadronic weak interactions. To lowest order in
the chiral-expansion, the effective chiral Lagrangian of the
Standard Model which describes the $\Delta S=1$ non-leptonic
interactions of on-shell pseudoscalar particles has only two
terms~:
$$\eqalignno{ {\cal L}^{\Delta S=1}_{eff} &=-{G_F\over
\sqrt{2}}V_{ud}V^{\ast}_{us}\left\{g_{\underline 8}\sum_i({\cal
L}_{\mu})_{2i}({\cal L}^{\mu})_{i3}+\right.\cr
&\left.g_{\underline{27}}\left[({\cal L}_{\mu})_{23}({\cal
L}^{\mu})_{11}+{2\over 3}({\cal L}_{\mu})_{21}({\cal
L}^{\mu})_{13}\right]+h.c.\vphantom{\sum_i}\right\}&(24) }$$
where ${\cal L}_{\mu}$ denotes the $3\times 3$ matrix
$${\cal L}_{\mu}=i f^2_{\pi}{\cal U}^{\dagger}D_{\mu}{\cal U}\ .
\eqno(25)$$ The two couplings in~(24) correspond to the effective
realization of the four-quark Hamiltonian which in the Standard
Model is obtained after integration of the heavy degrees of freedom
(i.e., the $t$-quark~; $W$ and $Z$~; $b$ and $c$ quarks) in the
presence of the strong interactions. The couplings proportional to
$g_{\underline 8}$ and $g_{\underline{27}}$ are the effective
realization of the four-quark operators which under chiral
rotations transform respectively as
$(8_L,\ 1_R)$ and $(27_L,\ 1_R)$. The strengths of the couplings
$g_{\underline 8}$ and $g_{\underline{27}}$, like $f_{\pi}$ and
$B$, are not fixed by symmetry requirements alone. Their values
can be extracted from $K\to\pi\pi$ decays with the results
$$|g_{\underline 8}|\simeq 5.1\ \hbox{and}\ g_{\underline
8}/g_{\underline{27}}\simeq 18.\eqno(26)$$
The huge ratio of these couplings shows clearly the enhancement
of the octet
$\Delta I={1/2}$ transitions. A direct calculation of these
couplings in the Standard Model has not yet been achieved. There
has been however progress in identifying the dynami\-cal sources
of this enhancement. (See refs~[26, 27] and references therein.)

Once the two coupling constants $g_{\underline 8}$ and
$g_{\underline{27}}$ are fixed , all  non-leptonic weak decays like
$K\to 3\pi$ and
$K\to\pi\pi\gamma$ are fully predicted to $O(p^2)$. There are also
a number of highly non-trivial $O(p^4)$ predictions for $K\to 3\pi$
decays which have been made~[28].

It has also been shown that non-leptonic radiative $K$-decays with
at most one-pion in the final state are forbidden to $O(p^2)$~[29].
This is due to the fact that electromagnetic gauge invariance
demands physical amplitudes to have a number of chiral powers
higher than just the two powers allowed by the lowest order
effective Lagrangian. Here the art of the game is to find physical
processes which are fully predicted at the chiral one-loop level,
like $K_S\to\gamma\gamma$~[30] and
$K_L\to\pi^{\circ}\gamma\gamma$~[33]~; or require only the
knowledge of a few number of $O(p^4)$ tree level couplings, like
$K^+\to \pi^+ e^+e^-$~; $K^+\to\pi^+\gamma\gamma$~;
$K_S\to\pi^{\circ}\gamma\gamma$. A very rich phenomenology has
been developed for these processes~[29, 32]. The study of the decay
$K_L\to\pi^{\circ}\gamma\gamma$ has also been particularly
helpful to clarify our views on the r\^ole of resonances in
$\chi$PT~[33]. The prospects to use $K_L\to \pi^{\circ} e^+e^-$ as a
possible new test of CP-violation in the Standard Model are
becoming rather good~[34].

\b

{\bf 6.} I mentioned in the introduction that the
$\chi$PT-approach has its own {\it limitations}. The basic problem
is the number of possible couplings with unknown coupling
constants which appear as we go to higher orders. In the sector of
the strong and electromagnetic interactions of pseudoscalars, the
VMD-approach I mentioned earlier can certainly help to fix some of
the new constants. However, the situation in the sector of the
non-leptonic weak interactions is rather dramatic. Here, the
number of possible
$O(p^4)$ couplings satisfying the appropriate $(8_L,\ 1_R)$ and
$(27_L,\ 1_R)$ transformation properties becomes huge [35, 36].
Only in the octet sector there already appear 35 independent terms
to describe on-shell pseudoscalar processes in the presence of
external fields. If we further restrict the external fields to only
photons, there are still 22 terms left. Applying the VMD-approach
does not help here because the primitive weak couplings of vector
and axial-vector particles to the pseudoscalars and among
themselves are phenomenologically unknown (See however
ref.~[37].)

\b

{\bf 7.} During the last few years there have been some
developments in estimating the low-energy coupling constants of
the effective chiral Lagrangian, using various approximations
rooted in the underlying QCD-theory. The battle-horse here is the
large-$N_c$ expansion~[38]. I already pointed out that in the large
$N_c$ limit~: $L_2=2L_1$. Further\-more, in this limit $L_4$ and
$L_6$ are non-leading~; (the constant $L_7$ has a peculiar
$O(N_c^2)$-behaviour due to the contribution of
$\eta'$-exchange~[8].) The actual values of the six constants~:
$L_1, L_3, L_5, L_7, L_9$ and $L_{10}$ which are leading $O(N_c)$
in the large $N_c$ limit of QCD, remain however unpredicted from
symmetry arguments alone.

There is a model of large $N_c$ QCD at intermediate energies
which has been recently elaborated~[39]~; the so called extended
Nambu Jona-Lasinio model (ENJL-model). This model has the merit
that it embodies practically all the previous attempts to ``derive"
an effective low-energy action from QCD in different limits.
However, like all the other models and / or attempts so far
suggested, it has the drawback that it does not confine, (at least in
the naive way that it has been formulated until now.)

The ENJL-model can be viewed as an approximation of large $N_c$
QCD, where the only new interaction terms retained after
integration of the high-frequency modes of the quark and gluon
fields down to a scale $\Lambda_{\chi}$ at which spontaneous
chiral symmetry breaking occurs, are those which can be cast in
the form of four-fermion operators. The parameters of the model
are then $\Lambda_{\chi}$ and the two coupling constants $G_S$
and $G_V$ of the two-possible (scalar-pseudoscalar) and
(vector-axial) four-fermion couplings. These two couplings can be
traded for the mass $M_Q$ of the constituent chiral quark, which
appears as a non-trivial solution to the gap equation involving
$G_S$
$$G^{-1}_S={M^2_Q\over
\Lambda^2_{\chi}}\Gamma\left(-1,\ {M^2_Q\over
\Lambda^2_{\chi}}\right)=\exp\left({-M^2_Q\over
\Lambda^2_{\chi}}\right)-{M^2_Q\over
\Lambda^2_{\chi}}\Gamma\left(0,\ {M^2_Q\over
\Lambda^2_{\chi}}\right)\ ;\eqno(27)$$
and the effective axial
coupling $g_A$ of the constituent chiral quarks to the
pseudoscalar Goldstone bosons
$$g_A={ 1\over  1+4G_V{M^2_Q\over
\Lambda^2_{\chi}}
\Gamma\left(0,\ {M^2_Q\over
\Lambda^2_{\chi}}\right) }\ \cdotp\eqno(28)$$
Here $\Gamma\left(n,\ \epsilon={M^2_Q\over
\Lambda^2_{\chi}}\right)$ denote incomplete gamma functions~:
$\Gamma(n,\
\epsilon)=\displaystyle\int^{\infty}_{\epsilon}{dz\over
z}e^{-z}z^n$. Following the standard procedure of introducing
auxiliary fields, one can rearrange the ENJL-Lagrangian in an
equivalent Lagrangian which is only quadratic in the quark fields.
The quark fields can then be integrated out using e.g., the heat
kernel expansion technique. The resulting effective Lagrangian is
one of the standard effective chiral Lagrangians discussed e.g. in
refs. [20] and [22], which describe the interactions of Goldstone
fields (the
$0^-$octet) with the lowest-lying resonance states $S(0^+), V(1^-)$
and $A(1^+)$. However, the coupling constants and masses in the
effective Lagrangian appear now as functions of only the three
input parameters : $M_Q,\Lambda_{\chi}$ and $g_A$. For example
$$f_{\pi}^2={N_c\over
16\pi^2}4M_Q^2g_A\Gamma(0,M^2_Q/\Lambda^2_{\chi})\eqno (29)$$
and
$$M^2_V={3\over 2}{\Lambda^2_{\chi}\over G_V}{1\over
\Gamma(0,M^2_Q/\Lambda^2_{\chi})}=6M^2_Q{g_A\over
1-g_A}\ \cdotp\eqno(30)$$
The explicit forms one gets for the six
$O(N_c)\ L_i$ constants are shown in Table 2. We also show in this
table the numerical results corresponding to the fit 1 discussed in
ref. [39] which corresponds to the set of input parameter values :
$$M_Q=265\ MeV,\quad \Lambda_{\chi}=1165\ MeV,\quad
g_A=0.61\ .\eqno (31)$$ The overall picture which emerges from
this simple model is quite remarquable. In principle, one can also
calculate any higher-$O(p^6)$ coupling which may become of
interest.

The calculation of the QCD vector, axial-vector, scalar and
pseudoscalar two-point functions at low and intermediate energies
within the ENJL-model, has been recently reported in ref.~[40]. The
calculations have been made to leading order in the
$1/N_c$-expansion, but to all orders in powers of momenta
$Q^2/\Lambda^2_{\chi}$. This opens the possibility of evaluating
nonfactorizable contributions to nonleptonic weak matrix
elements. The\break successful determination of the
$\pi^+-\pi^{\circ}$ electromagnetic mass-difference is an
en\-cou\-ra\-ging first test. Reference~[40] discusses also some
reasons why this simple ENJL-model of low-energy QCD works so
well. The fact that the model incorporates automatically many of
the dynamical constraints of short-distance QCD, is certainly one
of the reasons.

\b

{\bf 8.} It is clear that $\chi$PT has provided a revival of interest
on low energy hadron physics. The first step of showing that it is a
useful approach has successfully been made. The next challenge is
the improvement of the level of precision in the predictions~; and
their extension to nonleptonic weak-decays. This will require a
deeper development of understanding of the link between
$\chi$PT and QCD than just symmetry principles alone. The
challenge both at the theoretical and phenomenological levels is
open~; and also the motivation for new higher precision
experiments.

\vfill\eject

{\bf Table 1 :} Phenomenological values of the couplings $L_i$. The
third column shows the source used to extract this information~;
the coefficients
$\Gamma_i$ are those of eq. (18).

$$\vbox{
\def\tvi{\vrule height 12pt depth 5pt width 0pt}
\offinterlineskip\halign{
\tv\quad\hfill#\quad\tv&
\hfill#\quad\kern8pt\tv&
\quad\hfill#\hfill\quad\tv&
\quad\hfill#\quad\tv\cr
\noalign{\hrule}
$i$\hfill& \quad$L_i(M_{\rho})\times 10^3$\kern-8pt& Source&
$\Gamma_i$\hfill\cr
\noalign{\hrule} 1& $0.7\pm 0.5$& $K_{e4},\
\pi\pi\to\pi\pi$& 3/32\hphantom{4}\cr 2& $1.2\pm 0.4$&
$K_{e4},\ \pi\pi\to\pi\pi$& 3/16\hphantom{4}\cr 3& $-3.6\pm
1.3$& $K_{e4},\
\pi\pi\to\pi\pi$& 0\hphantom{144}\cr 4& $-0.3\pm 0.5$& Zweig
rule&1/8\hphantom{44}\cr 5& $1.4\pm 0.5$&
$F_K\ :\ F_{\pi}$& 3/8\hphantom{44}\cr 6& $-0.2\pm 0.3$& Zweig
rule& 11/144\cr 7& $-0.4\pm 0.2$& Gell-Mann--Okubo, $L_5,\
L_8$& 0\hphantom{144}\cr 8& $0.9\pm 0.3$&
$M_{K^0}-M_{K^+},\ L_5,\ (m_s-\hat m)\ :\ (m_d-m_u)$&
5/48\hphantom{4}\cr 9& $6.9\pm 0.7$&
$\left<r^2\right>^{\pi}_{em}$& 1/4\hphantom{44}\cr 10& $-5.5\pm
0.7$&
$\pi\to e\nu\gamma$& -1/4\hphantom{44}\cr
\noalign{\hrule}
}}$$

\b

{\bf Table 2 :} Couplings of $O(N_c)$ in the ENJL-model of
ref.~[39], with $g_A$ defined in eq.~(28)~;
$\Gamma_n\equiv\Gamma\left(n,\
M^2_Q/\Lambda^2_{\chi}\right)$. The second column gives the
results corresponding to the input parameter values in (31). The
third column gives the experimental values of Table 1.

$$\vbox{
\def\tvi{\vrule height 14pt depth 5pt width 0pt}
\offinterlineskip\halign{
\tv~~#\hfill~~\tv& ~~\hfill#~~\tv& ~~\hfill#~~\tv\cr
\noalign{\hrule}
\offinterlineskip\hfill ENJL-expression& Fit 1\hfill& Exp.\hfill\cr
\noalign{\hrule}
\offinterlineskip$f^2_{\pi}={N_c\over 16\pi^2}4M^2_Q
g_A\Gamma_0$& $(89 MeV)^2$& $(93 MeV)^2$\cr
$L_1={N_c\over 16\pi^2}{1\over
48}\left[\left(1-g^2_A\right)^2\Gamma_0+4
g^2_A\left(1-g^2_A\right)\Gamma_1+2g^4_A\Gamma_2\right]$&
$0.85$\quad& $0.7\pm 0.5$\ \cr
$L_2=2L_1$& $1.7$\hphantom{5}\quad& $1.2\pm 0.4$\ \cr
$L_3={-N_c\over 16\pi^2}{1\over
8}\left\{\left(1-g^2_A\right)^2\Gamma_0+4
g^2_A\left(1-g^2_A\right)\Gamma_1\right.$&
$-4.2$\hphantom{5}\quad&
$-3.6\pm 1.3$\ \cr
$\left.\hphantom{L_3=}-{2\over
3}g^4_A\left[2\Gamma_1-4\Gamma_2+{3\over
\Gamma_0}\left(\Gamma_0-\Gamma_1\right)^2\right]\right\}$&
&\cr
$L_5={N_c\over 16\pi^2}{1\over
4}g^3_A\left[\Gamma_0-\Gamma_1\right]$&
$1.6$\hphantom{5}\quad& $1.4\pm 0.5$\ \cr
$L_8={N_c\over 16\pi^2}{1\over 16}g^2_A\left[\Gamma_0-{2\over
3}\Gamma_1\right]$&
$0.8$\hphantom{5}\quad& $0.9\pm 0.3$\ \cr
$L_9={N_c\over 16\pi^2}{1\over
6}\left[\left(1-g^2_A\right)\Gamma_0+2g^2_A\Gamma_1\right]$&
$7.1$\hphantom{5}\quad& $6.9\pm 0.7$\ \cr
$L_{10}={-N_c\over 16\pi^2}{1\over
6}\left[\left(1-g^2_A\right)\Gamma_0+g^2_A\Gamma_1\right]$&
$-5.9$\hphantom{5}\quad& $-5.5\pm 0.7$\ {\vrule height 14pt
depth 10pt width 0pt}\cr
\noalign{\hrule}
}}$$

\vfill\eject

\n {\bf REFERENCES :}

\m

\parindent 1truecm

\item{\hbox to\parindent{\enskip \hfill[1] \hfill}} J.S.~Bell and
E.~de Rafael, Nucl. Phys. {\bf B 11} (1969) 611.

\item{\hbox to\parindent{\enskip \hfill[2] \hfill}} M.~Gell-Mann and
Y.~Ne'eman, {\it The Eightfold Way,} Frontiers in Physics, W.A.
Benjamin publs., (1964).

\item{\hbox to\parindent{\enskip \hfill[3] \hfill}} S.L.~Adler and
R.~Dashen, {\it Current Algebras and applications to Particle
Physics,} Frontiers in Physics, W.A.~Benjamin publs., (1968).

\item{\hbox to\parindent{\enskip \hfill[4] \hfill}} S.~Weinberg,
Phys. Rev. Lett. {\bf 18} (1967) 507.

\item{\hbox to\parindent{\enskip \hfill[5] \hfill}} W.A.~Bardeen,
Phys. Rev. {\bf 184} (1969) 1848.

\item{\hbox to\parindent{\enskip \hfill[6] \hfill}} J.~Wess and
B.~Zumino, Phys. Lett. {\bf 37 B} (1971) 95.

\item{\hbox to\parindent{\enskip \hfill[7] \hfill}} E.~Witten, Nucl.
Phys. {\bf B 223} (1983) 422.

\item{\hbox to\parindent{\enskip \hfill[8] \hfill}} J.~Gasser and
H.~Leutwyler, Ann. of Phys. (N.Y.) {\bf 158} (1984) 142~; Nucl.
Phys. {\bf B 250} (1985) 465, 517, 539.

\item{\hbox to\parindent{\enskip \hfill[9] \hfill}} J.~Bijnens,
G.~Ecker and J.~Gasser, {\it Introduction to Chiral Symmetry,} in
{\it The DA$\phi$NE Physics Handbook,} eds. L.~Maiani, G.~Pancheri
and N.~Paver (Frascati 1992), Vol.~I p. 107.

\item{\hbox to\parindent{\enskip \hfill[10] \hfill}} G.~Ecker, {\it
Chiral Perturbation Theory,} lectures at the 4$^{\hbox{\sevenrm
th}}$ Hellenic Summer School on Elementary Particle Physics,
Corfu, 1992. Preprint CERN-TH. 6660/92 and UW Th Ph - 1992-44.

\item{\hbox to\parindent{\enskip \hfill[11] \hfill}} A.~Pich, {\it
Introduction to Chiral Perturbation Theory,} lectures at the
V~Mexican School of Particles and Fields, Guanajuato, M\'exico
1992. Preprint CERN-TH. 6978/93.

\item{\hbox to\parindent{\enskip \hfill[12] \hfill}} H.~Leutwyler,
{\it Nonperturbative Methods,} Proc. XXVI International Conference
on High Energy Physics, Vol. I p. 185, Dallas, USA (1992).

\item{\hbox to\parindent{\enskip \hfill[13] \hfill}} J.~Gasser and
U.G.~Mei\ss ner, Phys. Lett. {\bf B 258} (1991) 219.

\item{\hbox to\parindent{\enskip \hfill[14] \hfill}} H.~Leutwyler
and M.~Roos, Z. Phys. {\bf C 25} (1984) 91.

\item{\hbox to\parindent{\enskip \hfill[15] \hfill}}  J.~Bijnens,
Nucl. Phys. {\bf B 337} (1990) 635.

\item{\hbox to\parindent{\enskip \hfill[16] \hfill}}
C.~Riggenbach, J.~Gasser, J.F.~Donoghue and B.R.~Holstein, Phys.
Rev. {\bf D 43} (1991) 127.

\item{\hbox to\parindent{\enskip \hfill[17] \hfill}}  L.~Rosselet
{\it et al.,} Phys. Rev. {\bf D 15} (1977) 574.

\item{\hbox to\parindent{\enskip \hfill[18] \hfill}}  S.~Bellucci,
J.~Gasser and M.E.~Sainio {\it (to be published)}

\item{\hbox to\parindent{\enskip \hfill[19] \hfill}}  H.~Marsiske,
{\it et al.,} (Cristal Ball Coll.), Phys. Rev. {\bf D 41} (1990) 3324.

\item{\hbox to\parindent{\enskip \hfill[20] \hfill}} G.~Ecker,
J.~Gasser, A.~Pich and E.~de Rafael, Nucl. Phys. {\bf B 321} (1989)
311.

\item{\hbox to\parindent{\enskip \hfill[21] \hfill}} J.F.~Donoghue,
L.~Ramirez and G.~Valencia, Phys. Rev. {\bf D 39} (1989) 1947.

\item{\hbox to\parindent{\enskip \hfill[22] \hfill}} G.~Ecker,
J.~Gasser, H.~Leutwyler, A.~Pich and E.~de Rafael, Phys. Lett. {\bf
B 223} (1989) 425.

\item{\hbox to\parindent{\enskip \hfill[23] \hfill}} Y.~Nambu and
J.J.~Sakurai, Phys. Rev. Lett. {\bf 8} (1962) 79.

\item{\hbox to\parindent{\enskip \hfill[24] \hfill}} E.~Pallante and
R.~Petronzio, Nucl. Phys. {\bf B 396} (1993) 205.

\item{\hbox to\parindent{\enskip \hfill[25] \hfill}} J.~Prades,
Preprint CPT-93/P.2871.

\item{\hbox to\parindent{\enskip \hfill[26] \hfill}} A.~Pich and
E.~de Rafael, Nucl. Phys. {\bf B 358} (1991) 311.

\item{\hbox to\parindent{\enskip \hfill[27] \hfill}} A.J.~Buras and
M.K.~Harlander, {\it A Top Quark Story~: Quark Mixing, CP
Viola\-tion and Rare Decays in the Standard Model,} Preprint
MPI-PAE/PTh 1/92, TUM-T31-25/92.

\item{\hbox to\parindent{\enskip \hfill[28] \hfill}} J.~Kambor,
J.F.~Donoghue, B.R.~Holstein, J.~Missimer and D.~Wyler, Phys. Rev.
Lett. {\bf 68} (1992) 1818.

\item{\hbox to\parindent{\enskip \hfill[29] \hfill}} G.~Ecker,
A.~Pich and E.~de Rafael, Nucl. Phys. {\bf B 291} (1987) 692.

\item{\hbox to\parindent{\enskip \hfill[30] \hfill}} G.~D'Ambrosio
and D.~Espriu, Phys. Lett. {\bf B 175} (1986) 237~; J.L.~Goity,
Z.~Phys. {\bf C 34} (1987) 341.

\item{\hbox to\parindent{\enskip \hfill[31] \hfill}} G.~Ecker,
A.~Pich and E.~de Rafael, Phys. Lett. {\bf B 189} (1987) 363.

\item{\hbox to\parindent{\enskip \hfill[32] \hfill}} G.~Ecker,
A.~Pich and E.~de Rafael, Nucl. Phys. {\bf B 303} (1988) 665.

\item{\hbox to\parindent{\enskip \hfill[33] \hfill}} G.~Ecker,
A.~Pich and E.~de Rafael, Phys. Lett. {\bf B 237} (1990) 481.

\item{\hbox to\parindent{\enskip \hfill[34] \hfill}} D.A.~Harris
{\it et al.,} Preprint EFI-93-49.

\item{\hbox to\parindent{\enskip \hfill[35] \hfill}} J.~Kambor,
J.~Missimer and D.~Wyler, Nucl. Phys. {\bf B 346} (1990) 17.

\item{\hbox to\parindent{\enskip \hfill[36] \hfill}}
G.~Esposito-Far\`ese, Z. Phys. {\bf C 50} (1991) 255.

\item{\hbox to\parindent{\enskip \hfill[37] \hfill}} G.~Ecker,
J.~Kambor and D.~Wyler, Nucl. Phys. {\bf B 394} (1993) 101.

\item{\hbox to\parindent{\enskip \hfill[38] \hfill}} G.'t~Hooft,
Nucl. Phys. {\bf B 72} (1974) 461.

\item{\hbox to\parindent{\enskip \hfill[39] \hfill}} J.~Bijnens,
Ch.~Bruno and E.~de Rafael, Nucl. Phys. {\bf B 390} (1993) 501.

\item{\hbox to\parindent{\enskip \hfill[40] \hfill}} J.~Bijnens,
E.~de Rafael and H.~Zheng, Preprint CERN-TH. 6924/93,
CPT-93/P.2917, Nordita-93/43 N.P.

\end